\documentstyle[aps,prl,epsf,twocolumn]{revtex}
\begin{document}
\def\bea{\begin{eqnarray}}
\def\eea{\end{eqnarray}}
\def\a{\alpha}
\def\d{\delta}
\def\p{\partial} 
\def\nn{\nonumber}
\def\r{\rho}
\def\rv{\bar{r}}
\def\la{\langle}
\def\ra{\rangle}
\def\e{\epsilon}
\def\o{\omega}
\def\n{\eta}
\def\g{\gamma}
\def\break#1{\pagebreak \vspace*{#1}}
\def\f{\frac}
\twocolumn[\hsize\textwidth\columnwidth\hsize\csname
@twocolumnfalse\endcsname
\draft
\title{Molecular Elasticity and the Geometric Phase}
\author{Joseph Samuel and Supurna Sinha}
\address{Raman Research Institute,
Bangalore 560080,India\\}
\date{\today}
\maketitle
\widetext
\begin{abstract}
We present a method for solving the Worm Like Chain (WLC) model for twisting
semiflexible polymers to any desired accuracy. 
We show that 
the WLC free energy is a periodic function of the applied twist
with period $4\pi$. 
We develop an analogy between WLC elasticity and the geometric phase of a 
spin $\frac{1}{2}$ system. 
These analogies are used to predict elastic
properties of twist-storing polymers. 
We graphically display the elastic
response of a single molecule to an applied torque. This
study is relevant to mechanical properties
of biopolymers like DNA. 
\end{abstract}

]
\narrowtext

Some long molecules are as stiff as needles, others
as flexible as thread.
The elasticity of
a stiff molecule is dominated by energy,
while the elasticity of 
a flexible molecule is dominated by its configurational entropy.
It has lately become 
feasible \cite{smith} to stretch, bend
and twist single molecules to study their elastic properties. 
The subject of this paper is the elasticity of semiflexible polymers
in which there is competition between
energetic and entropic effects. 
We consider a polymer which can bend as well as twist \cite{rabin}.
The flexibility of such a molecule is
characterized by two dimensionless parameters
$\alpha = L_{BP}/L_{TP}$ and $\beta = L/L_{BP}$, where $L$ is the
length of the molecule and 
$L_{BP}$ and $L_{TP}$ are the bend and twist persistence
lengths. For example DNA has a bend persistence length of about $53  nm$ and 
a twist persistence length of about $70  nm$. 

The main purpose of this letter is to draw attention to an experimentally
relevant topological 
subtlety which has not been discussed in previous theoretical treatments of 
the elasticity of twisting polymers. To appreciate the point, take a strip of
paper (a ribbon or belt will do as well) and tape one end of the strip to 
a table. Pull the strip taut by its other end and twist it by four 
half-turns ($4\pi$ rotation). If you now slacken the strip, you will find 
that it is possible, keeping the end fixed, to pass the strip around the 
end. Pulling the strip taut will reveal that the $4\pi$ twist has been 
released. This demonstration shows that a polymer can release twist,
{\it two turns at a time} by going around its end. 
A twist of $2\pi$ cannot be so released but can be transformed to $-2\pi$. This
of course, merely illustrates the well known mathematical fact that 
the rotation group is doubly connected\cite{saku}. We will now see that 
this mathematical fact has concrete experimental consequences for the elastic
properties of twist-storing polymers. Mentally replace the table with a
translation stage, the strip with DNA molecule with a micron sized 
magnetic bead at 
its free end and let the twisting
be done with a magnetic field.
If the DNA molecule is about $16\mu$ long, and not pulled
taut, it can release $4\pi$ worth of twist by passing around the bead. 
One will find that twisting the bead by $4\pi$ is
equivalent to not twisting it at all! Under the influence of thermal 
agitation, the molecule will  explore all configurations which it can 
reach by continuous deformation. As we have shown, a configuration with 
a $4\pi$ twist can be continuously deformed to a configuration with $0$ 
twist. It follows that the free energies for these two situations are the 
same.
If one were to measure the free
energy $G(\psi)$ of the molecule as a function of applied twist $\psi$
(by say measuring the torque-twist 
relation),
one would find that $G(\psi)$ is a periodic function of $\psi$ with period
$4\pi$. 
From the $4\pi$ periodicity of the free energy, it follows that
the torque-twist relation and other measurable elastic properties 
have the same periodicity.

The above discussion is even more relevant to theoretical models
of polymers which do not incorporate self avoidance. Such a 
polymer is a `phantom chain' \cite{Nelson} and can pass 
through itself.
In the absence of self avoidance, a polymer does not even 
need to pass around its end: it can pass {\it 
through} itself and so release twist {\it two turns at a time}. 
Indeed, such effects have been seen in recent single molecule 
experiments\cite{strick} and even earlier 
\cite{brown}:
the enzyme topoisomerase II converts a real DNA into a 
`phantom chain' \cite{strick,CELL} and in the presence of this enzyme (which 
plays a crucial role in 
replication), the DNA molecule releases twist 
{\it two turns at a time}\cite{strick,brown}.  The release of twist 
through bending modes (geometric untwisting) has also been 
discussed in \cite{gold}.

It is thus clear that 
in theoretical models which do not have self avoidance, the free 
energy is a periodic 
function of applied
twist with period $4\pi$. Bearing this in mind, we will now explore
the Worm Like Chain (WLC) \cite{kratky} model 
for twisting polymers, which ignores self avoidance. 
As was emphasized recently\cite{marko2} in a review of single molecule
experiments, 
``the precision of control and quantitative measurement and simple
interpretation of these experiments make 
detailed theoretical analyses appropriate''. 
While there has been some theoretical progress\cite{Nelson,Bouchiat}
on the WLC for twisting polymers, interest has been confined 
primarily to 
the high tension regime, which is theoretically more tractable because 
the molecule has only small perturbations about a linear configuration. 
Experiments of Bensimon \cite{Bensimon} have also explored this 
regime. In contrast the present letter theoretically explores the non 
linear regime of small forces, where the perturbative methods described in 
\cite{Nelson,Bouchiat} are inapplicable. 
In this regime we find that the WLC free energy has a $4\pi$ 
periodicity in contrast to the aperiodic free energy of the high tension 
regime.

We first describe the WLC 
model and summarize the existing theoretical analyses of this model. We then
point out a physically relevant mathematical subtlety without which the
solution of the WLC model is incorrect.
We then correctly solve the WLC model by a combination of 
analytical and numerical techniques
and elucidate some of the elastic properties that emerge from 
the model in graphical
form. A concluding 
discussion interprets the results.

{\it WLC Model:} The WLC model ignores self avoidance and views the polymer  
as
 a framed space curve ${\cal
C}=\{\vec x(s),{\hat e}_i(s)\}, i=1,2,3,$ $0\le s\le L$ of contour length 
$L$,
with an energy cost $\cal{E}({\cal C})$ for bending and twisting.
We
suppose
that one end of the polymer is tethered to the origin $\vec x (0) = 0$
and the other end at $\vec x (L) = \vec r$ is tagged. The unit tangent
vector ${\hat e}_3= {d{\vec x}/ds} $ to the curve describes the 
bending of the polymer while the twisting is captured by a
unit vector
${\hat e}_1$ normal to ${\hat e}_3$.
${\hat e}_2$ is then fixed by
${\hat
e}_2={\hat e}_3
\times {\hat e}_1$ to complete the right handed moving frame
${\hat e}_i(s), i = 1,2,3 $. 
The rate of change of 
the moving frame ${\hat e}_i(s)$ along the curve can be 
measured by its ``angular velocity vector'' ${\vec \Omega}$ defined by
\begin{equation}
\frac{d}{ds} {\hat e}_i(s)= {\vec \Omega}\times {\hat e}_i(s).
\label{angular}
\end{equation}
The components of ${\vec \Omega}$ in the moving frame
are $\Omega_i={\vec \Omega}.{\hat e}_i$ and the energy ${\cal E}({\cal C})$
of a
configuration ${\cal C}$ is given by
\begin{equation}
{\cal E}[{\cal C}]=1/2 \int_0^L[A ((\Omega_1)^2+(\Omega_2)^2)+
C (\Omega_3)^2]ds,
\label{energy}
\end{equation}
where $A$ is the bending modulus and $C$ the twist modulus.

Imagine that the ends of the polymer and the frames at these
ends $({\hat e}_i(0),{\hat e}_i(L))$ are held fixed. We wish
to compute the number of configurations

\begin{equation}
Q({\vec r},{\hat e}_i(0),{\hat e}_i(L))=\Sigma_{\cal C} \exp(-{\cal
E}[{\cal C}]/kT),
\label{sum}
\end{equation}
counted
with Boltzmann weight
$\exp(-{\cal E}[{\cal C}]/kT)$,  
which start at the origin with initial frame
${\hat e}_i(0)$ and end at ${\vec r}$ with final frame ${\hat e}_i(L)$.
The function $Q$ is related to the free energy of the molecule and
its measurable elastic properties like the force-extension relation
(FER)
and the torque-twist relation (TTR). An
overall multiplicative constant is not important in the calculation of
$Q$. This only leads to an additive constant in the free energy,
which drops out on differentiation and does not affect elastic
properties.

Let us fix a ``lab frame'' ${\tilde e}_i^b$ and write 
${e}^a_i(s)=R^a{}_b(s) {\tilde e}_i^{ b}$,
where $a=1,2,3$ is a vector index (as opposed to the frame index $i$) and
$R(s)\in SO(3)$ is a $3\times 3$ rotation matrix. 
There is a clear analogy between the elastic properties of the WLC and the 
motion of a top. Indeed as Mezard and Bouchiat\cite{Bouchiat} and Moroz
and Nelson\cite{Nelson} point out, the WLC problem can be mapped to the 
quantum mechanics of a symmetric top. From this mapping, one may
naively conclude \cite{footn} that the periodicity of the
free energy is $2\pi$. We now show that a careful treatment of the path
integral (\ref{sum}) gives the correct $4\pi$ periodicity.

Each configuration ${\cal
C}$ in (\ref{sum}) is characterized by a curve $\{R(s)\}$ in the rotation 
group with
fixed end points $R(0)$ and $R(L)$. The sum  
is over all configurations which are sampled under the influence of
thermal agitation. It is evident that we should only sum over
configurations $\{R(s)\}$ in a {\it single} homotopy class: thermal
agitation only causes continuous deformations and therefore 
cannot knock the polymer out of its homotopy class. 
$Q$ thus depends not only on ($\vec r, R(0),R(L)$) , but also on the
homotopy class  $[\{R(s)\}]$  of the paths 
$\{R(s)\}$ being summed over. This information is
exactly captured by going to the covering space $SU(2)$ of the rotation
group $SO(3)$. {\it This step is essential to correctly describe the 
elastic properties of the twisting polymer}. (If we do not take this 
step but remain on $SO(3)$, 
we are effectively summing over both homotopy classes,
which is a {\it physically incorrect} procedure.) 
The result is that
while the WLC {\it Hamiltonian} is the same as that of the top,
the WLC {\it configuration space} is not the configuration space $SO(3)$ 
of 
the top, but its double 
cover $SU(2)$ \cite{saku}. 
As we will see below, this
results in a $4\pi$ periodicity for the free energy.
$SU(2)$ is the same as $S^3$, the four dimensional sphere defined by
$\{x^\alpha,\alpha=1,2,3,4,\},\Sigma_\alpha (x^\alpha)^2=1$. In fact,
$x^\alpha$ are the Cayley-Klein parameters \cite{goldstein}
traditionally used in describing tops.

Let $\{g(s)\}$ be a 
continuous curve in $SU(2)$ which maps down to the curve $\{R(s)\}$.
$g(s)$ satisfies the differential equation 
$\frac{d{g(s)}}{ds}=i/2({\vec\Omega}.{\vec\sigma}) { g(s)},$
whose solution is a path ordered exponential: 
$g(s)=P[exp\int_0^s{ i{\vec\Omega(s')}.{\vec\sigma}/2 ds'} ] g(0).$
${\cal C}$ is now described by a curve $\{g(s\}$ in $SU(2)$,
with fixed end points $g(0)$ and $g(L)$. The standard Euler angles
$(\theta,\phi,\psi)$ on the rotation group can be used as co-ordinates
on $SU(2)=S^3$ if the range of $\psi$ is extended to $4\pi$ \cite{saku}. 
$SU(2)$ acts on itself by right and left action 
generated \cite{Landau} by space fixed
$(J_x,J_y,J_z)$ and body fixed $(J_1,J_2,J_3)$ angular momenta.

We can now write (\ref{sum}) more correctly 
as $Q(\vec r, q_0,q_L)$ where
$q_0=g(0)$ and
$q_L=g(L)$, to explicitly display the homotopy class dependence of $Q$. 
$Q(\vec r, q_0,q_L)$ has the path integral representation:
\begin{equation}
{\cal N}\int D[g(s)] e^{[-{{\cal E}({\cal
C})}/{k_BT}]} \delta(\vec x(L) - \vec r).
\label{path}
\end{equation}
$\cal N$ is a normalisation constant and the path integral is over all
paths
that go from $q_0$  to $q_L$ on $S^3$.
We now pass from $Q({\vec r},q_0,q_L)$ to its Laplace transform 
defined as ${\tilde Q}(f,q_0,q_L)=\int d{\vec r}exp[{\vec f}.{\vec
r}/L_{BP}]Q(\vec r, q_0,q_L)$,
where $L_{BP}=A/kT$. Performing the elementary integrations
and changing variables to $\tau=s/L_{BP}$ and 
${\vec \omega}={\vec \Omega}L_{BP}$, we see that 
${\tilde Q}(f,q_0,q_L)$ can be represented as ${\cal N}Z(f,q_0,q_L)$, 
where $Z$ has the path integral representation
\begin{equation}
\int D[g(\tau)] e^{-[\int_0^\beta d\tau
\frac{1}{2}(\omega_1^2+\omega_2^2+\alpha^{-1}\omega_3^2)-{\vec f}.
{\hat e_3}]} ,
\label{path1}
\end{equation}
where $\alpha=A/C=L_{BP}/L_{TP}$.
This is clearly the 
quantum amplitude 
$<q_L|exp[-\beta H_f]|q_0>$ for a particle on the surface of a $3$-sphere
to go from an initial 
position $q_0$ on $S^3$ to a final position $q_L$ in imaginary
time $\beta$ in the presence of an external force field. 
The Hamiltonian
is $H_f=H_0-f \cos{\theta}$, where
$H_0=1/2({J_1}^2+{J_2}^2+\alpha {J_3}^2)$, which is the Hamiltonian
of a symmetric top. If the exact eigenstates of $H_f$ were known,
we could write:
\begin{equation}
Z=\Sigma_n \exp[-\beta E_n] u_n^*(q_0) u_n(q_L),
\label{kernel}
\end{equation}
where $\{u_n(q)\}$ is a complete set of normalised eigenstates
of the Hamiltonian $H_f$ and $E_n$ are the
corresponding eigenvalues. 
Even though $H_f$ cannot be diagonalised analytically,
we can exploit its symmetries to reduce the problem to a
numerically tractable form.
$J_z$ and $J_3$ commute with the Hamiltonian $H_f$, 
reflecting the symmetry under space-fixed and body fixed
rotations about the third axis. As a result, $Z$ depends only
on the differences $\phi=\phi_L-\phi_0$ and $\psi=\psi_L-\psi_0$
and we write $Z(f,\theta_0,\theta_L,\phi,\psi)$.
Consider the dependence of $Z$ on $\psi$.
Since all the wave functions in (\ref{kernel}) are single valued
functions on $S^3$, it follows 
that $Z(\psi)$ is periodic in $\psi$ with period
$4\pi$: $Z(\psi+4\pi)=Z(\psi)$. This means that the free energy
$G=-1/\beta Log[Z]$ and all elastic properties have
the same period. 
This is {\it the first main 
result} of this letter. 
The fact that the periodicity is $4\pi$ and {\it not} $2\pi$ can be traced
to the fact the sum in (\ref{kernel}) extends not only over
tensorial states but also over spinorial ones. 
The variation of
$G$ with respect to the variables $(f,\theta_0,\theta_L,\phi,\psi)$
gives 
the elastic response to stretch
($f$),
bend ($\theta,\phi$) and twist ($\psi$). $G$ can be computed numerically
for any value of its arguments using mathematica 
programs \cite{math} that run for a few minutes on a PC,
using methods similar to those of \cite{bend}.  
From $G$ we can extract all possible information regarding
the elasticity of a polymer with bend and twist degrees of freedom. For 
instance one can predict the form of extension versus twist curves for
various values of the stretching force.

The $4\pi$ periodicity of the free energy strongly motivates the use
of spinorial methods. In fact, the WLC configuration space $S^3$ is the 
same
as the set of normalised states of a spin-$1/2$ quantum system. We will 
now show that there is a mapping between
a configuration of a twisting polymer and the quantum evolution of 
a spin-$1/2$ system. This brings out an interesting connection between 
$WLC$ elasticity and the 
Geometric Phase.
Let us introduce a $2$
component complex vector (a spinor) $\xi^1=x^1+ix^2,\xi^2=x^3+ix^4$,
which is normalised ($\xi^\dagger \xi=1$). We can write
$\xi^1=\cos{\theta/2}\exp^{-i\phi/2}\exp^{i\psi/2}$,
$\xi^2=\sin{\theta/2}\exp^{i\phi/2}\exp^{i\psi/2}$ and thus introduce
co-ordinates
$(\theta,\phi,\psi)$ ranging from $0$ to $(\pi,2\pi,4\pi)$
respectively. These are similar to Euler angles on the rotation group
and differ only in the range of $\psi$. The frame 
${\hat e}_i$ can be expressed 
as 
${\hat e}_3=\xi^\dagger {\vec \sigma} \xi$, 
${\hat e}_1+i{\hat e}_2=\xi^T(i\sigma_2) {\vec \sigma} \xi$,
where the $\sigma$s are the usual Pauli matrices. Notice that altering 
$\psi$ by $2\pi$ flips only the sign of $\xi$ and therefore does not 
affect the frame.  Using this mapping between $2$ component 
spinors and frames, we can import ideas from the geometric
phase to understand WLC elasticity. The information 
in the spinor $\xi$ can be decomposed into an overall phase $\psi/2$
describing twist and a ray ${\hat e}_3$ describing bend. Fix a
configuration ${\cal C}$ and note that $\xi(s)=g(s)\xi(0)$ satisfies
the Schr{\"o}dinger differential equation
$i\frac{d\xi(s)}{ds}={\hat h} \xi(s)$,
where ${\hat h}=-{\vec \Omega}.{\vec \sigma}/2$ is the 
``Hamiltonian'' of a spin half particle in an external magnetic
field ${\vec \Omega}$. 
We can now decompose the
difference $\psi/2=\psi(L)/2-\psi(0)/2$ between the final and initial
phases into a geometric phase and a dynamical phase.
The dynamical phase is given by the integral of the expectation
value of the ``Hamiltonian'' $\int_0^Lds\xi^{\dagger}{\hat h}\xi$,
which, using the definition of ${\hat e}_3$ above is seen to be
$\int_0^Lds
\Omega_3/2$,
half the twist $\Omega_3$ integrated along the polymer. The geometric
phase is given by half the solid angle swept out by the ray ${\hat
e}_3(s)$.
If the initial and final rays are distinct (but not antipodal),
one can join them by the unique shorter
geodesic \cite{Samuel} to enclose a solid angle.
The total twist difference $\psi$ is the sum of the ``dynamical
twist''-the integrated local twist-
$\int \Omega_3 ds$ and the ``geometric twist''- the solid angle swept
out by the tangent vector- which depends on the bending of the polymer.
In the literature\cite{Nelson,Bouchiat,maggs}, the distribution of applied
twist between twisting and bending is compared with the decomposition
of link into twist and writhe. This result is referred to as
White's theorem \cite{White} (though an earlier reference is
\cite{Cal}). The discussion \cite{Nelson,Bouchiat,maggs} applies to
closed, self-avoiding polymers which have been twisted an integral number
of times.
In contrast, our treatment applies also to
open polymers which have
been twisted a fractional number of times. However, since
the WLC model does not take into account self avoidance,
a twist of $4\pi$
is equivalent to no twist and the integral part of the twist
is only measured modulo two.
Our treatment captures the fractional 
part of the applied twist (which is geometrical) and the earlier
treatment captures its integral part (which is topological).
In this sense, the two discussions are complementary.
The analogy between twist elasticity and the geometric phase
is the {\it second main result} of this letter. The analogy has also
been noted in Ref\cite{maggs}, which, however, uses a vectorial 
correspondence 
rather than a spinorial one.
The decomposition of applied twist into a geometrical and a
dynamical part
leads to a coupling between the bend and the twist degrees
of freedom and has a direct bearing on the elastic properties 
of the WLC polymer.
As a specific
illustration we give the results for the special case of pure twist 
elasticity.  

{\it Pure twist elasticity:} 
We suppose that the tagged end is not constrained in position, but 
only in orientation. Integrating $Q({\vec r},q_0,q_L)$ over ${\vec r}$, we 
see that the applied 
force $f$ vanishes. We also suppose that the initial ${\hat e}_3(0)$ and
final ${\hat e}_3(L)$ tangent vectors are both in the  
same direction (which we take to be the $z$ direction). We compute the
distribution $Z(\psi)$ of $\psi$.
In this case only states for which $m=g$ contribute \cite{Edmonds} and 
$Z$ takes the form $Z=\Sigma_{g} {\cal Z}_{gg}$. Using
standard techniques from angular momentum theory, we can 
express $Z$ as $Z = \Sigma_{g}{e^{ig\psi}}{Q_g}$
where $Q_g =\Sigma_{j=|g|}^{\infty} (2j+1) exp[-\beta/2
(j(j+1)+(\alpha-1)g^2)]$, where $j$ runs in integer steps.
By an inverse Fourier transform, we compute $P(\psi)$ and the 
free energy $G(\psi)=-1/\beta Log[P(\psi)]$. By differentiating
with respect to $\psi$, we compute the torque 
$\Theta=\partial F/\partial \psi$ needed to twist the molecule
by an angle $\psi$. The torque-twist
relation is plotted in Fig.1. This graph, which is {\it the third main result} 
of
this letter, 
describes the pure twist elastic properties of a molecule in the WLC
model.
These graphs are easily interpreted in terms of the geometric
phase ideas described earlier. For large $\alpha$ ($\alpha$ is the
ratio of the bend to the twist persistence length), twist costs very
little energy, the molecule twists without bending, 
and as it takes hardly any torque to twist the molecule, the
TTR is almost flat. As $\alpha$ decreases, the applied twist
is shared between the twist and the bend. When $\alpha$ is zero
twisting is prohibitively expensive and
the  applied twist is all taken up by the bend. This causes the molecule
to buckle just as a towel does when it is wrung. When $\alpha=0$,
the ``polarisation vector'' ${\hat e}_1$ is parallel transported
along the polymer. The distribution $Z(\psi)$ then reduces to 
the distribution of solid angles (Berry Phases) enclosed by closed 
Brownian paths on the (Poincare) sphere,
which was calculated in \cite{BM} in the context of depolarised light
scattering.

In this letter we have solved the WLC model with bend and 
twist degrees of freedom and noticed analogies to spin$1/2$ systems and
the geometric phase.  
These analogies lead to 
a description in terms of a
particle on a sphere in 
external gravitational and magnetic fields. 
Such analogies, apart from giving us analytic tools to 
solve the problem virtually exactly for the first time, 
also provide
simple physical pictures: 
Imposing a twist on a molecule is like applying a magnetic field. 
The helical shape of a towel when it is wrung is similar to the
helical trajectory of a particle in a magnetic field. We hope 
this letter will encourage experimental work on twisting polymers
in the nonlinear low tension regime and set up 
a dialog between the theory and 
experiments on molecular elasticity.

{\it Acknowledgements:} It is a pleasure to thank A. Dhar, Y. Hatwalne
B. R. Iyer, V.A. Raghunathan and M. Rao for their comments
and R. Goldstein for his constructive suggestions which improved
the paper and for drawing attention to related work\cite{gold}.

\vbox{
\epsfxsize=4.0cm
\epsfysize=-0.8cm
\begin{figure}
\caption{The Torque-Twist relation for 
$\beta =L/L_{BP}=1$, $\alpha=L/L_{TP}=0,1,7$.}
\label{puretwist} 
\end{figure}}

\begin{references}
\bibitem{footn}
(see eq. 8 of \cite{Nelson})
This remark does not however, invalidate the main conclusions of
Refs.\cite{Nelson,Bouchiat}. These works are primarily interested in
the large force limit of the WLC. Release of $4\pi$ twist is 
then exponentially suppressed by energetics. The present study does not 
address this regime.
\end{references}
\end{document}